\documentclass[]{spie}  

\newcommand\arcsec{\mbox{$^{\prime\prime}$}}%
\newcommand\degree{\mbox{$^{\circ}$}}%

\let\OLDitemize\itemize
\renewcommand\itemize{\OLDitemize\addtolength{\itemsep}{-5pt}}

\usepackage{amsmath,amsfonts,amssymb}
\usepackage{graphicx}
\usepackage[colorlinks=true, allcolors=blue]{hyperref}
\usepackage[colorinlistoftodos]{todonotes}

\title{Expected observing efficiency of the Maunakea Spectroscopic Explorer (MSE)}

\author[a]{Nicolas Flagey}
\author[a]{Kei Szeto}
\author[a]{Kevin Ho}
\author[a]{Billy Mahoney}
\author[b]{Alan McConnachie}
\author[a]{Alexis Hill}
\author[a]{Calum Hervieu}

\affil[a]{Canada-France-Hawaii Telescope Corporation, 65-1238 Mamalahoa Hwy, Kamuela HI 96743, USA}
\affil[b]{NRC Herzberg, Dominion Astrophysical Observatory, 5071 West Saanich Road, Victoria, British Columbia, Canada}

\authorinfo{Further author information: (Send correspondence to N.F.)\\N.F.: E-mail: flagey@cfht.hawaii.edu}

\begin{document} 
\maketitle

\begin{abstract}
The Maunakea Spectroscopic Explorer (MSE) will obtain millions of spectra each year in the optical to near-infrared, at low ($R\simeq 3,000$) to high ($R\simeq 40,000$) spectral resolution by observing $>$4,000 spectra per pointing via a highly multiplexed fiber-fed system. Key science programs for MSE include black hole reverberation mapping, stellar population analysis of faint galaxies at high redshift, and sub-km/s velocity accuracy for stellar astrophysics.

One key metric of the success of MSE will be its survey speed, i.e. how many spectra of good signal-to-noise ratio will MSE be able to obtain every night and every year. This is defined at the higher level by the observing efficiency of the observatory and should be at least 80\%, as indicated in the Science Requirements.

In this paper we present the observing efficiency budget developed for MSE based on historical data at the Canada-France-Hawaii Telescope and other Maunakea Observatories. We describe the typical sequence of events at night to help us compute the observing efficiency and how we envision to optimize it to meet the science requirements.
\end{abstract}

\keywords{observing efficiency, fiber, spectrograph, MSE}

\section{INTRODUCTION}
\label{sec:intro}  

The Maunakea Spectroscopic Explorer (MSE) is a project to upgrade the 3.6-meter telescope and instrumentation of the Canada-France-Hawaii Telescope (CFHT) to a 11.25-meter telescope equipped with fiber-fed spectrographs dedicated to optical and near-infrared (NIR) spectroscopic surveys. The current baseline for MSE is that of a prime focus, 10-meter effective aperture telescope feeding a bank of low (LR) and moderate (MR) spectral resolution spectrographs (LR, R$\sim$3000 and MR, R$\sim$6000) located on  platforms, as well as high (HR) spectral resolution spectrographs (HR, R$\sim$20000-40000) located on the more stable pier of the telescope. The 1.5 square degree field of view of MSE will be populated with 3249 fibers of 1\arcsec\ diameter allocated to the Low and Moderate Resoluton (LMR) spectrographs, and 1083 fibers of 0.8\arcsec\ for the HR spectrographs. The fibers will all be positioned with spines from the Sphinx system, with both LMR and HR fibers being available at all time.

At the previous SPIE Astronomical Telescopes and Instrumentation meeting, the status and progress of the project were detailed in Ref.~\citenum{Murowinski2016} while an overview of the project design was given in Ref.~\citenum{Szeto2016} and the science based requirements were explained in Ref.~\citenum{McConnachie2016}. An update of the project at the end of conceptual design phase is presented this year in Ref.~\citenum{Szeto2018a} with a review of the instrumentation suite in Ref.~\citenum{Szeto2018b}. Other papers related to MSE are focusing on: the summit facility upgrade (Ref.~\citenum{Bauman2016, Bauman2018}), the telescope optical designs for MSE (Ref.~\citenum{Saunders2016}), the telescope structure design (Ref.~\citenum{Murga2018}), the design for the high-resolution (Ref.~\citenum{Zhang2016, Zhang2018}) and the low/moderate-resolution spectrograph (Ref.~\citenum{Caillier2018}, the top end assembly (Ref.~\citenum{Mignot2018, Hill2018b}), the fiber bundle system (Ref.~\citenum{Venn2018, Erickson2018}), the fiber positioners system (Ref.~\citenum{Smedley2018}), the systems budgets architecture and development (Ref.~\citenum{Mignot2016, Hill2018}), the observatory software (Ref.~\citenum{Vermeulen2016}), the spectral calibration (Ref.~\citenum{Flagey2016a, McConnachie2018a}), the throughput optimization (Ref.~\citenum{Flagey2016b, McConnachie2018b}), the injection efficiency (Ref.~\citenum{Flagey2018c}), and the overall operations of the facility (Ref.~\citenum{Flagey2018a}).

In this paper we focus on the observing efficiency of the facility, which is a science requirement for MSE, at 80\%. Given the design choices for MSE, and using information provided during the Conceptual Design Phase, a systems budget was established and a model was developed to estimate the expected observing efficiency. The plan of the paper is as follows. In section \ref{sec:defin} we define the observing efficiency, in section \ref{sec:sys} we detail the budget, and in section \ref{sec:mod} we present how the budget was used to compute the observing efficiency.

\section{Defining the observing efficiency}
\label{sec:defin}  

The observing efficiency is defined as the fraction of time the observatory is collecting photons divided by the time the observatory could have been collecting photons, which is all the time available for observations except that lost to weather.

\begin{eqnarray}
	\label{eq:oe_def}
	OE = & \frac{\rm{nighttime~spent~collecting~photons}}{\rm{all~nighttime}-\rm{nighttime~lost~to~weather}} \\
    = & \frac{ \rm{all~nighttime}-\rm{nighttime~lost~to~weather}-\rm{nighttime~lost~to~other~factor} }{\rm{all~nighttime}-\rm{nighttime~lost~to~weather}}
\end{eqnarray}

The observing efficiency is defined in “steady state” operations for MSE, i.e. after commissioning of the observatory. In addition, we assume it is averaged at least over a year, given the nature of the typical events occurring at a ground based astronomical facility.

\subsection{Night duration}
\label{sec:night}  

Using CFHT’s QSOTools Almanac tab for the entire year of 2016, we established that a night is between 10.35 and 12.75 hours long. However, the usable nighttime depends on the instrument in use and on how far below the horizon the Sun has to be. Typically, an IR instrument can be used for longer than an optical instrument, though in J and H bands the gain might be limited. At CFHT, the rule for the optical imager Megacam is to use it between 12\degree\ twilight while the NIR imager Wircam can be used between 8\degree\ twilight. The optical spectrograph Espadons is also used between 8\degree\ twilight. On average, the 8\degree\ night has a duration of 10.8$\, \pm \,$0.84 hours while the 12\degree\ night has a duration of 10.2$\, \pm \,$0.86 hours. MSE will use both LMR and HR spectrographs simultaneously. Therefore, MSE will be able to operate at full capacity only after 12\degree\ twilight. It is possible however that MSE will obtain science data outside of the 12\degree\ twilight. We thus assume that, on average, a night will be 10.2 hours long.

\subsection{Weather losses}
\label{sec:wea}

The impact of weather has to be accounted for in order to calculate the observing efficiency. In addition, an understanding of the historical weather conditions at CFHT site is helpful to define the typical “observing conditions” (e.g. humidity, wind speed, temperature) during which MSE will be collecting photons from astronomical targets.

Using the QSO Tools for CFHT, which logs information about time losses, among other, we found that over the course of 2014, 2015, and 2016, CFHT lost a total of 1090, 1024, and 834 hours due to weather. This corresponds to an average loss of 2.7 hours per night attributed to weather. Over a longer period of time (2007-2016), the average time lost to weather is 794 hours per year, or 2.2 hours per night, which corresponds to an average of 21\% of a typical night. This search also highlights the typical variations of losses due to weather from year to year. As a more striking illustration of the weather losses variations, we looked at the total losses from January 1st 2018 to May 1st 2018 and found that 777 hours, which corresponds to 6.5 hours per night, had already been lost to bad weather.

We then use the information provided in the QSO Tools by the CFHT observer to define observing conditions. Parsing all comments, the words that most often came up are “humidity” (43\%), “fog” (27\%), “cloud” (23\%), “snow” (16\%), “ice” (13\%), “precipitation” (13\%), and “wind” (11\%). All of the historical data for CFHT do not systematically record cloud coverage and their impact on weather losses. The word “cloud” however appears in 18\% of the weather loss comments without being combined with the words “humidity” or “wind”. The words “humidity” and “wind” are cited for 51\% of the time lost to weather, thus 10.5\% of the total time. We therefore look for a realistic combination of wind and humidity conditions that corresponds to about 10\% of the time and use it to define “observing conditions”.

\begin{table}[h]
\centering
\caption{\label{tab:cfht_weather} Statistics about the weather at the CFHT site, between the local time of 19:00 and 05:00, between the year 2007 and 2016.}
	\begin{tabular}{| l | c c c c |}
		\hline
        Percentile & 80 & 90 & 95 & 99 \\
        \hline
        Humidity (\%) below& 58 & 85 & 96 & 100 \\
        Wind speed (knots) below & 20.5 & 27.5 & 34.5 & 51.5 \\
        Temperature (\degree C) within & -2.55 to 4.35 & -3.55 to +5.15 & -4.35 to +5.85 & -6.15 to 7.35 \\
        1-hour temperature variation (\degree C) below & 0.65 & 1.05 & 1.45 & 2.85 \\
        \hline
	\end{tabular}
\end{table}

The QSO Tools for CFHT records weather data (temperature, wind speed, and humidity) every hour. We decide to limit our analysis to the nighttime hours (between 19:00HST and 5:00HST). Some statistical information about the weather data are summarized in Table \ref{tab:cfht_weather}, including temperature and temperature variations, which have no direct impact on weather losses but may be relevant to subsystems design teams. We also look at combinations of weather data and find that:
\begin{itemize}
\item 19\% of the time, either wind OR humidity was outside of their 90\%-ile,
\item 11\% of the time, either wind OR humidity was outside of their 95\%-ile.
\end{itemize}
However, the 95\%-ile for humidity is at 96\% which is unsafe for observing conditions due to risk of condensation. We thus limit the humidity to its 90\%-ile and find that:
\begin{itemize}
\item 14\% of the time, either wind is outside its 95\%-ile OR humidity is outside its 90\%-ile,
\item 11\% of the time, either wind is outside its 99\%-ile OR humidity is outside its 90\%-ile.
\end{itemize}

This last combination corresponds to conditions with wind speed below 51.5 knots and humidity below 85\% and leads to time losses that are very close to the 10\% weather loss due to wind and humidity that we are looking for. They also match the CFHT weather policies: closing the dome when the wind is above 50 knots or humidity is above 85\%.

In the budget for observing efficiency, we then assume that weather losses account for 2.2 hours per night, with about half of it due to wind or humidity. This leaves 8.0 hours per night for operations, which is the denominator in equation \ref{eq:oe_def} and most of the nominator. The residual in the nominator is derived from the budget (see section \ref{sec:sys}).

\subsection{Other observatories}

Before presenting the budget for the MSE observing efficiency, and to provide additional context, we discuss the observing efficiency at three other facilities: CFHT, DESI, and Gemini.

\subsubsection{CFHT}

Over the course of the 2014, 2015, and 2016 years, CFHT lost a total of 1274, 1281, and 987 hours of observing time, including 184, 257, and 153 hours {\bf not} due to weather. Over 10 years (2007-2016), the average losses are 1032 hours per year, including 237 hours per year {\bf not} due to weather. This corresponds to an average loss of 0.65 hour per night (very similar to those allocated for MSE: 0.66 hour per night, see section \ref{sec:mod}). However, these losses do not account for time spent not collecting photons because, e.g., the telescope is slewing or the detectors are being readout.

The most important contributors to the 0.65 hours per night can vary quite significantly from year to year. For instance, over the 2014-2016 period, the main contributors are “engineering time” (277 hours in 3 years, 47\%), “MegaCam” (98 hours, 17\%), “queue scheduling” (48 hours, 8\%), and “focusing” (47 hours, 8\%), which is part of normal operations. Over the 10-year period however, the main contributors are “dome shutter failure” (830 hours in 10 years, 35\%), “engineering time” (592 hours, 25\%), “MegaCam” (247 hours, 10\%), and “queue scheduling” (226 hours, 10\%). Therefore, while several items were always significantly contributing to time losses, particular events can have devastating effects on the observing efficiency.

To account for the other contributors that impact the observing efficiency and are not accounted for in the time losses of CFHT’s QSOTools (i.e. overheads including readout, telescope slew, filter change, calibrations), we used the CFHT database to find how many hours each instrument (MegaCam, Wircam, Espadons, Sitelle) was actually collecting photons, by summing all exposure times (see Table \ref{tab:oe_cfht_1}). We add all science exposures graded well enough for science analysis (1-3 in the CFHT grading scale). It is important to note here that CFHT’s PI put constraints on the observational conditions for the observations which complicates the scheduling but should not impact the overall observing efficiency. We combine these numbers with an average night length of 10.5 hours to derive the observing efficiency at CFHT. Between 2010 and 2016, the observing efficiency at CFHT was about 54\%, with extremes at 45\% and 63\%. Scaled to all time available including that lost to weather, the observing efficiency is 42\% on average.

\begin{table}[t]
	\centering
    \caption{\label{tab:oe_cfht_1} Statistics on instrument use at CFHT in number of hours spent collecting photons.}
	\begin{tabular}{| c | c c c c c c c |}
		\hline
		Year &  2010 &  2011 &  2012 &  2013 &  2014 &  2015 &  2016 \\
		\hline
        Espadons           & 531.9 & 340.3 & 632.7 & 638.8 & 701.8 & 731.1 & 755.4 \\
		MegaCam            & 858.3 & 528.3 & 545.6 & 774.6 & 631.3 & 750.2 & 750.7 \\
		Sitelle            &     - &     - &     - &     - &     - &  15.7 &   165 \\
		WirCam             & 346.0 & 479.4 & 295.1 & 163.6 & 178.1 & 152.2 & 222.3 \\
        \hline
		Total              &  1736 &  1348 &  1473 &  1577 &  1511 &  1649 &  1893 \\
		\hline
		Observing efficiency & 50.4\% & 46.8\% & 45.1\% & 57.7\% & 55.1\% & 58.7\% & 62.9\% \\
        \hline
	\end{tabular}
\end{table}

Weather losses and technical losses are summarized in Table \ref{tab:oe_cfht_1} and account on average for 22\% and 7\% of the whole time available, but both vary significantly from year to year: 10 to 29\% for weather losses, 3 to 18\% for technical losses. Scaled to the time not lost to weather, to be comparable to the observing efficiency, all technical losses account on average for 9\% of observing efficiency losses ($7\times(1-0.22)$). From these numbers we can infer the fraction of nighttime spent pointing the telescope, obtaining calibrations, reading out the detectors, and other processes not collecting science photons: 29\% of all time ($100-22-42-7$) or 37\% of the time not lost to weather ($29/(1-0.22)$).

\begin{table}[t]
	\centering
    \caption{\label{tab:oe_cfht_2} Statistics about weather losses and technical losses at CFHT, given in number of hours and as fraction of the total time available.}
	\begin{tabular}{| c | c c c c c c c |}
		\hline
		Year   &   2010 &   2011 &   2012 &   2013 &   2014 &   2015 &   2016 \\
		\hline
		Weather losses       &    390 &    953 &    574 &   1099 &   1090 &   1024 &    833 \\
		relative to total time & 10.2\% & 24.9\% & 14.9\% & 28.7\% & 28.4\% & 26.7\% & 21.7\% \\
		\hline
		Technical losses     &    121 &    159 &    692 &    301 &    184 &    257 &    153 \\
		relative to total time &  3.2\% &  4.2\% & 18.0\% &  7.9\% &  4.8\% &  6.7\% &  4.0\% \\
		\hline
	\end{tabular}
\end{table}

\subsubsection{DESI}

In its Conceptual Design Report\footnote{\url{http://desi.lbl.gov/wp-content/uploads/2014/04/DESI_CDR_20140827_1135.pdf}}, the DESI experiment expects the following:

\begin{quote}
	\textit{We project that 57\% of the scheduled time will deliver usable data, where “usable data" is assumed in conditions when the dome is open and seeing is better than 1.5 arcsec. Although DESI will observe when the seeing is worse than 1.5 arcsec, those data have been ignored in these estimates of survey duration.}
\end{quote}

While this is not exactly the same definition of observing efficiency as we have for MSE, in particular because of the seeing limit at 1.5 arcsec, it gives an interesting point of comparison, especially from a facility with highly multiplexed optical spectrograph. The typical delivered image quality at the Mayall Telescope has a median of 1.1\arcsec\ and it seems like 1.5\arcsec\ roughly corresponds to the 90\%-ile of the image quality distribution (see e.g. Ref.~\citenum{Dey2014}). Converting their observing efficiency to the MSE definition (i.e. not limited by image quality) would thus lead to a value of about 63\%, significantly worse than MSE’s requirement. No further comparison is possible without knowing all the details about how the DESI experiment obtained their number.

\subsubsection{Gemini}

The observing statistics at both Gemini North and South from 2005 to 2012 are available publicly\footnote{\url{http://www.gemini.edu/sciops/statistics}}. These numbers reveal that between 46 and 75\% of the time was spent on science and between 12 and 37\% was lost to weather, which corresponds to observing efficiencies that range from 63 to 91\%, with a mean of 79\%. However, this is not equivalent to the definition of the observing efficiency we use for MSE, as it seems to include all overheads typically part of the science operations (e.g. telescope slew, readouts, calibrations).

In 2005, after about a year of science operations, a presentation was given on the observing efficiency at Gemini\footnote{Presentation on observing efficiency at \url{http://www.gemini.edu/science/GSC/GSC2005-10/}.}, reporting on the "open shutter" time, and their definition was stated as:
\begin{quote}
	Sum of all science exposures plus calibrations obtained between evening and morning nautical twilight divided by the usable time available. [...] Usable time obtained from hours between nautical twilights less time lost
due to weather and technical faults.
\end{quote}

This is still not equivalent to our definition, as they include time spent on calibration exposures. They reported average values per instrument between 59 and 70\% with peak efficiency of 78 to 88\% and a slightly better efficiency at Gemini North than Gemini South.

The comparison with a few other observatories leads to typical observing efficiency (MSE's definition) of about 60\%. Aiming at 80\% for MSE, as motivated by the Science Requirements, therefore appears challenging. However, our goal is to demonstrate hereafter that a facility dedicated to spectroscopic survey (i.e. not a multi instrument observatory), whose systems are designed with reliable components, operated in an efficient way, and supported by a staff of great experience, can achieve such observing efficiency.

\section{Systems budget}
\label{sec:sys}  

In this section we account for all observing time losses except weather. We first present the typical nighttime sequence of events and describe how we account for partial downtimes. We then detail the budget for observing efficiency.

\subsection{Nighttime sequence}
\label{sec:nightseq}

\begin{figure}
\includegraphics[width=\linewidth]{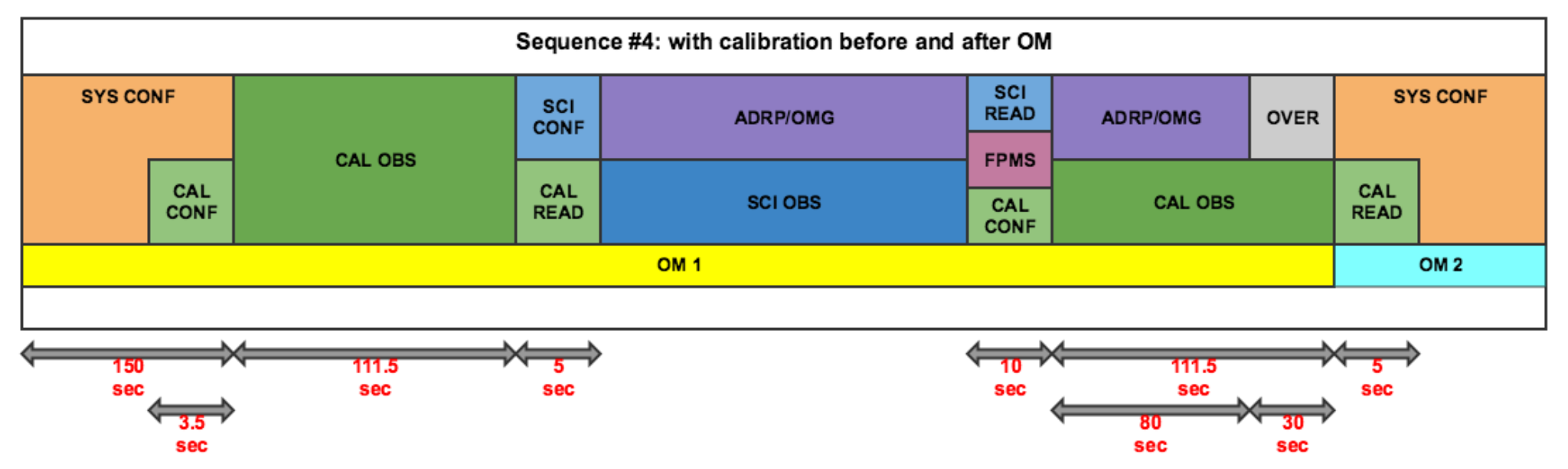}
\caption{\label{fig:seq} Nominal nighttime sequence of events for MSE. The time allocated for each "block" is indicated below the sequence.}
\end{figure}

The nominal sequence of events is described in the Operations Concept for MSE and in Figure \ref{fig:seq}. Hereafter are the definitions for the "blocks" used in the sequence.
\begin{description}
\item[SYS CONF:] system configuration (telescope slew, enclosure rotation, positioners move, …)
\item[SCI OBS:] science observations (collecting photons from a given set of science targets with all readouts except the last)
\item[SCI READ:] last science readout of the SCI OBS
\item[FPMS:] accurate measurement of the fibers positions after SCI OBS by the fiber positioner metrology system (FPMS)
\item[CAL CONF:] calibration configuration (turning on calibration unit)
\item[CAL OBS:] calibration observations (collecting photons from calibration unit, not including turning it on/off and last readout)
\item[CAL READ:] calibration readout (last readout of the CAL OBS, turning off the calibration unit)
\item[SCI CONF:] science configuration (usually after a CAL OBS to make sure the guiding is still good for science observations)
\item[ADRP/OMG:] automatic data reduction pipeline (ADRP) and updating schedule via the observing matrix generator (OMG)
\item[OVER:] MSE Staff override of the schedule update
\end{description}

In the sequence, we assume the following processes occur in parallel:
\begin{description}
\item[SYS CONF, CAL CONF:] turning the calibration unit on can happen while the system is being configured.
\item[CAL READ, SCI CONF:] while the last calibration exposure is readout and the calibration unit turned off, the calibration system is turned on, the system will verify guiding is still active.
\item[SCI OBS, ADRP/OMG:] while the science exposures are being obtained and readout, the automatic data reduction pipeline will run and provide real-time feedback to the scheduler.
\item[CAL CONF, SCI READ, FPMS:] while the last science exposure is readout, the calibration unit can be turned on, and the fiber metrology system can measure the position of the fibers at the end of the science observations.
\item[CAL OBS, ADRP/OMG, OVER:] while the calibration exposures are being obtained and readout, the automatic data reduction pipeline will finish processing the previous science observations run and provide real-time feedback to the scheduler.
\item[SYS CONF, CAL READ:] while the last calibration exposure is readout and the calibration system is turned off, the system can start configure for the next SCI OBS block.
\end{description}

Different sequences other than in Figure \ref{fig:seq} have been envisioned, with different order in which the CAL OBS and ADRP/OMG are executed. The baseline for MSE is however to use the sequence with calibration exposures obtained right before and right after the science observations, for improved data reduction and calibration. Once the system is well characterized, a decision might be made to switch to a different sequence, with higher observing efficiency.

\subsection{Accounting for partial downtime}
\label{sec:partial}  

MSE will be a complex facility, with more than 4000 robotic positioners, connected via tens of meters of optical fibers to several units of LMR and HR spectrographs. Some events will prevent the entire system from collecting science photons (e.g. issue with the aperture on the enclosure) and under these circumstances, the time spent not collecting science photons is fully deducted from the observing efficiency. Other events will prevent part of the system from being operational, though it will still collect science photons. For instance, if only one out of four spectrographs is not working for 1 hour at night while the other three are collecting photons, only 15 minutes are removed from the observing efficiency.

\subsection{System configuration}
\label{sec:sys_conf}

\begin{figure}
\centering
\includegraphics[width=.75\linewidth]{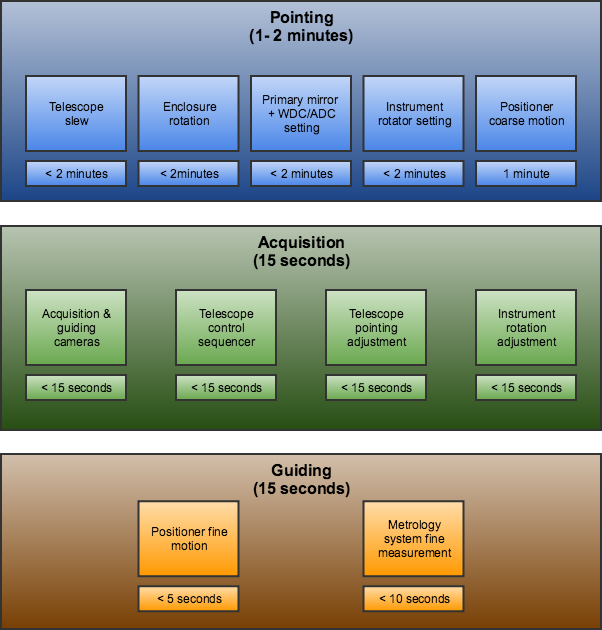}
\caption{\label{fig:config} Sequence of events during Pointing, Acquisition, and Guiding. During Pointing and Acquisition, all indicated processes can occur in parallel. During Guiding, both processes occur in sequence. The spectrographs will be configured in parallel with all three blocks of Pointing, Acquisition, and Guiding and just need to provide back-illumination for the fiber positioner metrology system.}
\end{figure}

Before each new observation, the whole system will need to be configured (e.g. the telescope will slew, the enclosure will rotate, the positioners will move the fibers). Some subsystems configurations will occur in parallel, others will occur in sequence (see Figure \ref{fig:config}).

In our budget, the system configuration will last at most 2.5 minutes. The exact breakdown for the system configuration depends on the actual sequence of observations during the night. For instance, if a given observation is repeated, the telescope and enclosure will not need to slew, and the positioners may not need to be reconfigured. However, we can list hereafter all the possible constituents of a system configuration.

First, we assume the typical configuration sequence is broken into Pointing, Acquisition, and Guiding. The Pointing procedure is basically the coarse, “open-loop” positioning of most subsystems, based on an assumed astrometric solution from lookup tables, using sky to focal surface coordinates mapping. We allocate 2 minutes total to the Pointing procedure, which consists of the following subsystems operating in parallel:
\begin{itemize}
	\item Telescope slew (2 minutes): the telescope may have to move from one side of the sky to the opposite. Slew range will be limited - in frequency and amplitude - via efficient scheduling, but it could be as large as 180\degree. Typical azimuthal slew rates are of the order of one to a few degrees per second (3\degree/s for the Discovery Channel Telescope, 1.3\degree/s for Keck). Elevation slew rate will not be as critical since most targets will be observed at high elevation (ZD50 and above). In the Telescope Structure DRD, the slewing requirement is to be faster than a trapezoid profile with acceleration/deceleration of 0.2\degree/s$^2$ and top speed of 1.5\degree/s in azimuth, and 0.2\degree/s$^2$ and 1.0\degree/s in elevation. This corresponds to a 180\degree slew in less than 130 seconds in azimuth and 60\degree in less than 70 seconds in elevation. Hence, we allocate 120 seconds to the telescope slew.
	\item Primary mirror (2 minutes): a complete shape and optical axis orientation adjustment may need to be performed each time the telescope changes pointing. This procedure will need to be done before the telescope has reached its expected position and will nominally rely on lookup tables.
	\item Enclosure slew (2 minutes): the enclosure should be in position before the telescope is, so that the acquisition process can start right away. During conceptual design phase, the slewing requirement was to be faster than a trapezoid profile with acceleration/deceleration of 0.25\degree/s$^2$ and top speed of 1.25\degree/s. This corresponds to a 180\degree\ slew in less than 150 seconds. We decide to allocate 120 seconds as we will surely avoid 180\degree\ slews.
	\item Instrument rotator, hexapod and WFC/ADC (2 minutes): the instrument rotator (InRo), hexapod, and WFC/ADC should be in position before the telescope is, so that the acquisition process can start right away and the fibers positioners metrology system can fine-tune the fibers positions.
    \item Positioners coarse motion (1 minute): the coarse motions will occur while the telescope, hexapod, and instrument rotator are in action. The fine motion occurs later.
\end{itemize}

It is important to note that some operations will be much faster if the next SCI.OBS block is not too far in the sky from the previous (e.g. telescope slew and enclosure rotation), which is one of the goals of the OMG. However, the Pointing procedure will surely never be shorter than 1 minute because some other operations always require the same amount of time regardless of sky position (e.g. positioners coarse motion).

The Pointing procedure is followed by the Acquisition procedure that corresponds to the fine, closed-loop positioning of most subsystems. Based on the results of the Pointing procedure, i.e. the location of guiding stars on the Acquisition and Guiding Cameras (AGC), the astrometric solution (accurate sky to focal surface coordinates mapping) and errors will be computed, and offset commands sent to multiple subsystems. At CFHT, the acquisition procedures last between 14 seconds for Wircam, 40 seconds for Megacam, and 90 seconds for Espadons. For MSE, the acquisition process will be complex, involving the telescope, hexapod, WFC/ADC, instrument rotator, but this will benefit from the anticipated improvements in computational power and analysis techniques. In addition, acquisition will need to be automatized and not require human intervention. For MSE, we allocate 15 seconds to the Acquisition procedure which includes the following subsystems operating in parallel:
\begin{itemize}
	\item AGC (15 seconds): the exposure on the acquisition and guiding cameras will be read. This is expected to be a fast process since the detectors will not be very large and the frames should be able to tolerate higher readout noise given the magnitude of the stars used for guiding.
    \item Control sequencer, telescope pointing, and InRo adjustment (15 seconds): the astrometric solution will need to be computed and commands sent to the control sequencers. Ideally one pointing offset command will be sufficient to reach the pointing accuracy required to start Guiding. Likewise, one angular offset command will be sufficient to reach the accuracy required to start Guiding. During Conceptual Design Phase, the InRo speed, assuming observing accuracy, can be up to 15 degrees per minute. This converts into about 1.3 mm per second at the edge of the InRo, which in turn converts into an on-sky angular speed of about 13 arcsec per second. In just one second then, the InRo can make an accurate correction to its angular position that is significantly larger than the telescope pointing accuracy.
\end{itemize}

Once the Acquisition procedure is complete, Guiding can start. This means guide stars have been put at the desired locations on the AGC and an accurate sky to focal surface coordinates mapping has been achieved. The last step in the positioners’ configuration will occur only after this is done. Once Guiding is effective, the following subsystems will operate in sequence  (15 seconds) right before the science observation:

\begin{itemize}
	\item Positioners fine motion (5 seconds): the fine motions require the instrument rotator to be in place, so the metrology system can accurately measure the fibers’ positions. According to the Conceptual Design of the PosS/FPMS, the entire process of positioning fibers will take between 30 and 70 seconds. However, the fine motion by itself is listed at 0.3 to 2.8 seconds. Adding some buffer for overheads leads to a 5-second allocation.
    \item FPMS fine measurement (10 seconds): at the end of the positioners’ fine motion, the FPMS will be used to accurately measure the position of each fiber. In the Sphinx CoDR report, 10 seconds are allocated to this step.
\end{itemize}

During the positioners fine motion, it is important to note that Guiding needs to be “ON” while the FPMS is in use, so an accurate astrometric solution is known and not lost. This means that the AGC shall be able to operate to maintain guiding while the back-illuminated fibers and fiducials in the focal surface emit light.

In parallel to the Pointing, Acquisition, and Guiding procedure, the spectrograph will have plenty of time to be configured as it will usually be limited to switching from low- to medium-resolution in the LMR units. While the PosS/FPMS are in use, the spectrographs provide back-illumination through the fiber train. We expect the spectrograph setup not to be a time limiting factor. The only requirement is that the operations within the spectrographs (e.g. change of mode between LR and MR) work in parallel without interfering with other operations. The whole 2.5 minutes are available for spectrograph setup. We note that while the whole system is being configured, the spectrographs may be reading out the most recent exposures (science or calibration, see sections 4.5 and 4.4), which will shorten the time available for spectrograph setup. We do not foresee this to be an issue within the allocated 2.5 minutes. 

\subsection{Calibrations}
\label{sec:sys_cal}

Nighttime calibrations are expected before and after each SCI OBS block. While the hardware design to provide those calibration exposures is not defined yet, we provide a baseline allocation for the expected procedure. There might be two different sets of calibration exposures to obtain at night: flat and arcs. There might also be a set of calibrations for LMR observations and another for HR observations. We baseline multiple exposures for both flats and arcs, to mitigate issues that could occur on a single exposure (e.g. cosmic rays), although we ultimately aim to only require a single exposure. With a baseline of 3 arcs and 3 flats for each set of calibrations, we need to allocate time for a total of 12 calibration exposures.

Nighttime calibration exposures will use at most 4 minutes, broken down as two blocks: one before and one after each SCI OBS block. The calibration time will be the sum of the time spent collecting photons, reading out the detectors, and turning the calibration system on/off. Nighttime calibrations will use the same observatory configuration as during the science observations: the telescope will be pointing at the same field of view, the enclosure's aperture will be open, the positioners will be allocated to the same targets, and the spectrographs will be using the same mode.

\begin{itemize}
	\item Readouts (18 seconds): detectors read-out with low noise (a few electrons) can occur at a frequency of about 1 MHz (e2v 231 series, 6k by 6k, 3 MHz max, 5e- at 1 MHz, 2e- at 50 kHz). For calibration exposures, low read-out noise is not necessary and the fastest read-out rate will be used (12 seconds). Binning (2x2 for HR, 2x1 for LMR) will shorten the readout time to 6 seconds, and using all 4 outputs will decrease it to 1.5 second. The total allocated time for all readouts is thus 18 seconds.
	\item Turning on/off the calibration system (3.5 seconds each): this includes moving any mechanical part of the calibration system (e.g. deploying a screen). Some of this time will be spent in parallel with other processes.
	\item Collecting calibration photons (95 seconds): to limit the time spent on each calibration block to a reasonable allocation (2 minutes total), we have 95 seconds left to allocate to the time spent collecting calibration photons.
\end{itemize}

\subsection{Science observations last readout}
\label{sec:sys_read}

Most readouts of science observation exposures are included in the time allocated to the SCI OBS block, as described in section \ref{sec:nightseq} of the current document. The time for the last readout however needs to be accounted for separately, because the system is no longer collecting photons, and therefore this process can occur in parallel with other processes. The time allocated for a science observation readout is 10 seconds.

All science readouts will use a low-noise mode, since science data will be collected. The e2v detectors from the 231 series (6k by 6k) can be read between 50 kHz with a noise of 2 electrons and 1 MHz with a noise of 5 electrons. Our baseline here is for an intermediate frequency of 500 kHz, which corresponds to a readout time of 72 seconds. This readout time can be reduced to 36 seconds by binning pixels (2x2 for HR, 2x1 for LMR), and 4 outputs shorten it to 9 seconds.

\subsection{Fiber metrology system (end of observations)}
\label{sec:sys_fpms}

At the end of a science observation, we will use the fiber metrology system (FPMS) to accurately measure the position of the fibers. This information will be useful for calibration. It will also help to confirm the final position of the fibers if open-loop repositioning is used in the future.

We allocate 10 seconds to this use of the FPMS, based on the Sphinx Conceptual Design report  configuration time, where a final accurate measurement of the fibers is allocated 10 seconds when positioning the fibers before the science observations. We will use this exact same procedure at the end of the science observation to measure the amplitude of the motion of the positioners.

\subsection{Realtime feedback}
\label{sec:sys_rt}

After each exposure in a SCI OBS block, the automatic data reduction pipeline (ADRP) is executed to provide realtime feedback so the schedule can be adjusted if necessary. The time constrain arises at the end of a SCI OBS block, when all exposures have been readout: the data reduction and decision to update the schedule will need to occur fast enough after a SCI OBS block is completed but before the system is configured for the next SCI OBS.

Since these operations will occur in parallel with the calibration exposures that are obtained right after the SCI OBS block, they will need to occur in less than the 2 minutes allocated for the calibrations minus the 10 seconds allocated for the last science readout (see section \ref{sec:sys_read}). In addition, in order to allow for human intervention to override the decision to update the schedule (at least in the early phase of the project), to which we allocate 30 seconds, we end up with 1 minute and 20 seconds for the data reduction and schedule update processes. It is difficult to estimate which process will be the more complex: reducing many exposures and estimating SNR for a significant amount of targets, or using this information to optimize a field with thousands of targets. We therefore allocate 40 seconds to both processes.

In summary, the allocation is broken down as follows:
\begin{itemize}
\item Reduction of data: 40 seconds.
\item Update of schedule: 40 seconds,
\item Time for a human intervention: 30 seconds.
\end{itemize}

The goal for MSE is to always have the schedule up-to-date and to have the best SCI OBS block currently queued for observations. The schedule will “continuously” be updated throughout the night, after each exposure has been readout and when the data reduction has been completed.

\subsection{Primary mirror realignment and phasing}
\label{sec:sys_m1}

Every time segments from the primary mirror are exchanged during the day, a procedure to align and phase the primary mirror is required at night.

As described in the Operations Concept Document for MSE\footnote{Project internal document available upon request.}, we expect to change every segment of the primary mirror once every 2 years, to maintain its overall reflectivity. Based on the experience at Keck, we expect the MSE staff will be able to exchange 3 segments in a day using a similar procedure. As a consequence, there will be 10 phasing procedures per year. The other 2 months in the year, only a maintenance procedure will take place.

Feasibility studies of a phasing system for MSE suggest a 2-hour phasing procedure and a 0.5-hour maintenance procedure. Both procedures require nighttime, and even though they will use fairly bright stars (magnitude lower than 10), good contrast and atmosphere stability will be critical. We thus allocate the whole 21 hours to nighttime as a baseline. 

In summary, the phasing of the primary mirror will not use more than 21 hours per year:
\begin{itemize}
\item 2 hours per night, 1 night per month, 10 months per year, following a segment exchange, for a total of 20 hours per year
\item 0.5 hour per night, 1 night per month, 2 months per year, to maintain alignment, for a total of 1 hour per year
\end{itemize}

\subsection{Failure of subsystems}

In the following subsection, we discuss the time lost due to failures of subsystems. Not all failures impact the observing efficiency the same way. For instance, if one spectrograph, one positioner, or one fiber fails at night, science photons will not be collected for that spectrograph, positioner, or fiber only. If software fails, it could mean the control sequencers are not working and then no science is possible at all, or it could mean that the data reduction pipeline is failing, which does not prevent MSE from collecting photons. In each of the following subsections, we assume a night length of 8 hours as failures that occur when the weather prevents observations will not impact observing efficiency.

\subsubsection{Failure of positioners (PosS) and metrology system (FPMS)}
\label{sec:sys_poss}

To some extent, a positioner that cannot be properly allocated could still collect useful photons from the sky.  When a positioner fails, it could be for different reasons: (1) the positioner could simply be unable to move, (2) its initial allocation could not be accurate enough but remain stable during the science observations, or (3) its initial allocation could be accurate but move during the science observations. Using a conservative approach, we consider the spectrum collected on the spectrographs’ detectors are useless. Hereafter we consider that any positioner that cannot be properly allocated as a failure.

The instrument FMOS at Subaru, a PosS delivered by AAO, did not lose any positioner during science operations, though 8 out of 400 spines (2\%) were “lost” before the beginning of science operations. In addition, a total of 19 lost hours were attributed to FMOS (mostly related software), out of 259 nights of science observations (0.7\%). Scaling this up to the number of nights available after weather loss at CFHT (284 nights per year) leads to about 20 hours. Adding the 2\% of lost positioners before commissioning would lead to a total of 80 hours lost per year. However, MSE will request spares and will replace failed positioners at commissioning. Therefore, we assume that during commissioning, all positioners will be tested and replaced if needed to provide a PosS with 100\% of positioners functional. We thus account here only for positioners that would fail after commissioning (i.e. 20 hours).

In the AAO PosS requirements document for Conceptual Design, the primary performance requirement states that at most 2\% of the positioners will not meet all performance requirements (goal: 0\%), which does not mean these spines cannot be used. These 2\% include inoperative spines at delivery and permanent or temporary inoperative spines during their 20 years of operations. More specifically, the AAO design specifies that {\it "less than 1\% (goal: 0.1\%) of the spines shall fail during operational lifetime of the PosS, with a spine being considered to be failing when it cannot hit its targets (within defined tolerances) for more than 2\% of observations."} Taking a conservative approach where 1\% of the spines fail 100\% of the time, we allocate 30 hours observing efficiency loss to PosS.

We conservatively allocate to the fiber positioner metrology system (FPMS) the same observing efficiency loss of 30 hours per year. 

\subsubsection{Failure of fibers bundle system (FiBS)}
\label{sec:sys_fits}

A fiber is considered failing if it cannot collect photons and send them to the spectrograph. Fibers will prevent photons collected at the focal surface to reach the spectrograph if they are broken. Other issues could be related to wrongly placed fibers in the input slit of the spectrograph or to wrongly placed fibers in the positioner at the focal surface, but these issues are related to the integration phase, and not the fiber bundles.

A broken fiber will easily be detected, as no signal will be measured on the detectors for such fibers. Other more subtle issues could be related to fibers not delivering the signal as expected (e.g. larger than expected beam, inefficient scrambling). We consider all these issues “a failure”. We assume there will be spare fibers available. Given how complex it will be to access the top end assembly, especially near the positioners, we do not envision that every single failed fiber will be changed immediately. Here we assume that as long as 99\% of the fibers are providing signal as expected (i.e. less than 43 broken fibers), no repair will occur.

Therefore, the worst-case scenario is to operate with 1\% missing fibers, which corresponds to 1\% observing efficiency loss, i.e. 30 hours per year. We assume that the failing fibers are not associated with the 2\% failing positioners, to remain conservative.

\subsubsection{Failure of spectrographs}
\label{sec:sys_spec}

The failure of a spectrograph means that a fraction of the photons entering fibers at the focal surface will not be converted into science data on a detector. The remaining spectrographs will still collect data for science.

We use CFHT historical data as a comparison: over the past 10 years (2007-2016), time was lost because of Megacam (246.7 hours), Wircam (73.1 hours), and Espadons (16.7 hours). This lost time is related to failure of the instrument, not engineering tests and upgrades. In total, it corresponds to 336 hours of lost time over 10 years, or about 4 nights per year on average, due to instrument failure. However, time lost for imagers (Megacam and Wircam) corresponds to about 4-5\% of the total time spent collecting photons with those instruments, while for the spectrograph (Espadons) it is about 10 times lower.

For the MSE spectrographs, each treated as individual instruments, we scale the Espadons numbers to a year of full usage and thus allocate 10 hours per year and per spectrograph. In addition, each spectrograph only corresponds to a fraction of the targets in the focal surface and both HR and LMR units will be used at all time. Therefore, once summed, the total losses due to all spectrographs are 10 hours per year.

According to the LMR and HR Conceptual Design Phase reports, each of the 6 LMR spectrograph unit will carry about 550 fibers and both HR unit will carry 578 fibers. These numbers are sufficiently close for us to allocate the same fraction of the observing efficiency to each unit (about 1/8), and hence 75\% of 10 hours observing efficiency losses to all the LMR spectrographs and 25\% to all the HR spectrographs.

\subsubsection{Failure of other subsystems}
\label{sec:sys_other}

Several subsystems may prevent MSE from collecting any photons in the event of some particular failure. For instance, if the telescope or enclosure cannot be moved at all, science operations will not be possible. If the calibration unit is not working, MSE will still collect science photons and the impact on observing efficiency will be more limited. Hereafter we allocate time in terms of observing efficiency losses. We describe, when necessary, how to convert failure time into observing efficiency losses.

The combined losses of observing efficiency due to failure of subsystems shall not exceed 69 hours, allocated as follows:
\begin{itemize}
\item telescope structure (5 hours per year): from Conceptual Design Phase document on the telescope structure design.
\item enclosure (5 hours per year): from Conceptual Design Phase document on the enclosure design.
\item primary mirror (1 hour per year): the observing efficiency losses will be related to missing segments, and to otherwise failure of the whole system to operate.

Our baseline is to always observe with 60 segments, and spare segments should allow this. If a segment is not usable for science observations, it might be for multiple reasons. For instance, the crane might stop working during a segment exchange procedure, or the actuators/electronics might be failing during the night. For each missing segment, we consider that the observing efficiency is reduced by $\sqrt{1/60}$, i.e. less than 1\%.

We assume that one of the worst-case scenario is when a segment exchange procedure is interrupted for some reason. Three “old” segments have then been taken out but the freshly coated ones have not been put back. MSE will thus operate for a whole night with 57 segments instead of 60. An equivalent scenario would be to operate at night with three segments that have failing actuators. We assume that such a situation will not happen more than once per year. It corresponds to an observing loss of 0.4 hour per year ($3/60\times8$). In an extreme case, a dysfunctional segment handling system could lead to a broken segment, which would affect the observing efficiency, to the extent that the six similar segments in M1 would stay on the mirror cell and would not be exchanged for recoating.

Other failures that might prevent nighttime observations could be related to a failing alignment of the M1 segments during the night, though it would need to be major to prevent observations. In addition, any long-term failure of the coating facility could delay segment exchange to a point that M1 reflectivity does not meet requirements anymore, which could be translated into loss of observing efficiency. Our expectations are this will be extremely rare and we round up our allocation to 1 hour per year of observing efficiency loss due to M1.

\item wide field corrector / atmospheric dispersion corrector (1 hour per year): for this subsystem to prevent observations, it would need to be in a configuration that is not stable and does not allow the positioners to be allocated accordingly. In that case, the WFC/ADC mechanism could be disabled. If the WFC/ADC simply fails to reach the required configuration, e.g. because we disabled it, the FPMS and PosS shall be able to account for the error, in part, and though the injection efficiency, and thus the throughput, may be affected, it will surely remain insignificant and will not prevent observations. We therefore expect the WFC/ADC to extremely rarely impact observing efficiency and allocate 1 hour per year to this subsystem.

\item prime focus hexapod system (1 hour per year): similar to the WFC/ADC, for the PFHS to prevent observations, it would need to be unstable and move during an observation. As with the WFC/ADC, disabling the PFHS should be the first response so as to not decrease observing efficiency. Throughput might be affected however. As for the WFC/ADC, we allocate 1 hour per year of observing efficiency loss to this subsystem.

\item instrument rotator (5 hours per year): for the InRo to prevent observations, it simply needs to stop working as expected. If the InRo stops rotating, basically no observations will be possible, except maybe close to the keyhole. If the InRo rotates too slow or too fast, it may affect injection efficiency and thus throughput to such a point as to prevent observations. We allocate 5 hours per year of observing efficiency loss to this subsystem.

\item observing building facility (10 hours per year): the OBF could prevent observations for various reasons, including but not limited to, cooling issues, electricity issues, and heat dissipation issues. Those issues will either affect a given subsystem or a group of subsystems. Therefore, they may prevent the observatory to function at 100\% or to function at all. We allocate 10 hours per year of observing efficiency loss to this subsystem.

\item telescope optical feedback system (10 hours per year): Without optical feedback (i.e. acquisition and guide cameras) the telescope will not be able to guide and thus no science will possible. Indeed, without guiding, we will not be able to position the fibers accurately because no astrometric solution can be computed during the Pointing and Acquisition phases. We allocate 10 hours per year of observing efficiency loss to this subsystem.

\item science calibration unit (1 hour per year): if the science calibration unit (SCal) subsystem fails, this will not directly prevent MSE from collecting photons. At best, it will delay the execution of the calibration exposures, the data reduction, and maybe even the data distribution, but the science value of the observations will not be completely lost. In the case of MSE, once the system is well characterized, using calibration exposures obtained at a different time will likely not seriously impact the scientific return of the data.

At worst, the calibration exposures obtained after the SCal subsystem has been fixed will not be the “best” for reducing the science data obtained when SCal was not working, and the quality of the reduced data may be impacted as a consequence. It will not, however, make the science data completely useless.

While the impact of a failed SCal might be limited, we ideally want the SCal system to work every night (and day for daytime calibrations). We allocate a failure rate of 20 hours per year to SCal. The impact on Observing Efficiency, however, will be significantly smaller and we allocate 1 hour per year of observing efficiency loss to SCal.

\item software (30 hours per year): At CFHT, issues associated with the QSO Tools were responsible for 226 hours of lost time between 2007 and 2016. For MSE, if the scheduler (i.e. OMG) is not working, it will prevent observations as it will not generate SCI OBS blocks and send commands to the observatory. If the ADRP is not working, it will simply delay the data reduction and should not prevent the OMG to schedule and execute observations. Following on the history of CFHT, and assuming we will improve the reliability of the OMG with respect to the QSO Tools, we allocate 20 hours per year of observing efficiency loss this subsystem.

The other component of the software architecture, which comprises control sequencers for e.g. the telescope and the instruments, may prevent observations if any of the major control sequencers stop working. For instance, the telescope control sequencer at CFHT was responsible for 56 hours of observing losses in the past 10 years. In addition, part of the time losses charged to some instrument might also be associated with the control sequencers. As a result, we allocate an additional 10 hours per year of observing efficiency loss to this subsystem.

\end{itemize}

\subsection{Engineering on-sky time}

Engineering on-sky time is allocated to test and update subsystems, usually from a software point of view (e.g. telescope control sequencer). In order to minimize the nighttime losses due to on-sky engineering tests, simulators will be available to test changes and updates off the telescope before implementing them on the telescope. It is however envisioned that some on-sky engineering time will be required.

At Gemini, between 2005 and 2012, on average a few percent of the nighttime ($< 5\%$) every semester was allocated to engineering. At Keck I and II, 31 nights are allocated to telescope engineering, and 40 to instrument engineering, out of 730 available nights, which corresponds to almost 10\% of the year. Typically, CFHT allocates 7 nights of engineering time per semester. From 2014 to 2016, on-sky engineering operations have used 277 hours, for an average of about 92 hours (11.5 nights) per year, and 592 hours from 2007 to 2016, for an average of about 59 hours (7.5 nights) per year over that longer period.We then use these two different periods to highlight the variations in the demands for engineering on-sky tests.

From 2014 to 2016, the on-sky engineering losses were associated to various topics, summarized in Table \ref{tab:cfht_eng}. Some of the entries could actually be considered calibration of the observatory or instrument. Once folded into the observing efficiency, it actually does not matter how the time was lost.

\begin{table}
\centering
\caption{\label{tab:cfht_eng} Summary of the on-sky engineering use of the CFHT telescope over the past 3 and 10 years, in hours.}
	\begin{tabular}{| l | c c |}
        \hline
		Time period                                       & 2014-2016 & 2007-2016 \\
        \hline
        Total                                             &       277 &       592 \\
        \hline
        SITELLE instrument                                &        90 &        90 \\
        Telescope Control System                          &        90 &        90 \\
        Optical Turbulence Project                        &        18 &       110 \\
        Megacam instrument                                &        12 &        18 \\
        Wircam instrument                                 &         0 &        18 \\
        First Night of Run’s tests                        &        30 &        44 \\
        “Focus”, “Calibrations”, “Pointing”, and “Guider” &        20 &        89 \\
		Other                                             &        30 &       182 \\
        - staring mode                                    &         0 &        24 \\
        - cross talk                                      &         1 &        20 \\
        - OT Idle                                         &         1 &        38 \\
        \hline
	\end{tabular}
\end{table}

At MSE, we do not expect to have as much instrument development as at Keck, and we will follow and improve on the good engineering practices at CFHT. We use the CFHT 10-year statistics (60 hours/year) as our starting point. Then, we assume the amount of time allocated for engineering on-sky test of instruments and TCS can be decreased, at least after MSE’s first few years of operations, thanks to an increased use of off-sky simulators. In addition, multiple MSE subsystems will allow for tests while science operations are being executed. For instance, engineering tests on one LMR spectrograph (out of 6) or one HR spectrograph (out of 3) would only prevent a fraction of the fibers to be allocated to science targets. We consider that updating lookup tables for pointing and focusing models will be part of on-sky engineering time. We allocate 50 hours per year to engineering on-sky test.

\section{Model}
\label{sec:mod}  

We use the system budget detailed above to derive an average observing efficiency for MSE.

First, given the average length of a night (10.2 hours, see section \ref{sec:night}) and the typical losses to inclement weather (2.2 hours, see section \ref{sec:wea}), we derive an average usable nighttime of 8.0 hours.

We then subtract the time allocated for:
\begin{itemize}
  \item engineering on-sky tests (50 hours per year or 0.14 hour per night)
  \item primary mirror phasing alignment (21 hours per year or 0.06 hour per night)
  \item failure of subsystems (total of 169 hours per year or 0.46 hour per night) due to:
  \begin{itemize}
    \item telescope structure (5 hours per year)
    \item primary mirror (1 hour per year)
    \item wide field corrector and atmospheric dispersion corrector (1 hour per year)
    \item prime focus hexapod system (1 hour per year)
    \item instrument rotator (5 hours per year)
    \item enclosure (5 hours per year)
    \item observatory building and facility (10 hours per year)
    \item software (30 hours per year)
    \item calibration unit (1 hour per year)
    \item spectrographs (10 hours per year)
    \item positioners and metrology system (60 hours per year)
    \item fiber bundle (30 hours per year)
  \end{itemize}
\end{itemize}
 
This leads to 7.37 hours per night available for science, including all overheads. We then use the nighttime sequence (see section \ref{sec:nightseq} and Figure \ref{fig:seq}) to infer that the total overheads per SCI OBS block are 388 seconds:

\begin{itemize}
  \item system configuration: 150 seconds (see section \ref{sec:sys_conf})
  \item calibration: two times 111.5 seconds (120 seconds minus the two times 3.5 seconds to turn on or off the calibration unit and the 1.5 second for the least readout that can be spent in parallel with other processes, see section \ref{sec:sys_cal})
  \item science configuration: 5 seconds (see section \ref{sec:nightseq})
  \item last science readout: 10 seconds (see section \ref{sec:sys_read})
  \item data reduction, schedule update, and human intervention: 110 seconds (see section \ref{sec:sys_rt}).
\end{itemize}

To derive the total time required to observe a SCI OBS with overheads requires to know the duration of the SCI OBS block (see Table \ref{tab:oe}). Given that duration, the average number of SCI OBS block per night can be computed and thus the average nighttime spent collecting photons. For an average SCI OBS block duration of 30 minutes, we find that 12.1 SCI OBS block can be executed on average per night for an open-shutter collecting science photons time of 6.1 hours and 1.3 hours of overheads, for a final observing efficiency of 75.5\%, not far from the science requirement. Switching to a 1-hour average duration of the SCI OBS block leads to only 6.7 blocks being executed each night but an overall observing efficiency of 82.9\%. To just reach the 80\% requirements corresponds to an average SCI OBS length of 44 minutes. These durations, ranging from 30 to 60 minutes, are in line with what is currently envisioned from an operational point of view: MSE will reach for the faintest regions of the Universe and is expected to perform long SCI OBS block as often as possible.

\begin{table}[h]
\centering
\caption{\label{tab:oe} Overheads cost and overall observing efficiency for MSE given an average duration of the SCI OBS block.}
\begin{tabular}{| c | c c c |}
\hline
Total overheads per SCI OBS (seconds)             & \multicolumn{3}{c|}{388} \\
Average science time per SCI OBS (minutes)        &   30 &   44 &   60 \\
Average time per SCI OBS with overheads (seconds) & 2188 & 3028 & 3988 \\
\hline
Average number of SCI OBS per night               & 12.1 &  8.8 &  6.7 \\
\hline
Time spent on overheads (h/n)                     & 1.31 & 0.94 & 0.72 \\
Science time (h/n)                                & 6.06 & 6.42 & 6.65 \\
\hline
Observing efficiency (\%)                         & 75.5 & 80.0 & 82.9 \\
\hline
\end{tabular}
\end{table}

\section{Conclusions}

In this paper we have presented a detailed budget for the observing efficiency of MSE, relying on (1) a wealth of historical data collected at CFHT over the past decade, as well as other observatories on Maunakea and elsewhere, (2) on engineering information provided during conceptual design phase by the various subsystems design teams, and (3) when none of the above were available, on educated guesses from support staff with decades of experience in the field at Maunakea observatories. This has allowed us to show that it is possible to meet the science requirement of an observing efficiency at 80\% for MSE, pending observing blocks of $> 44$ minutes of science observations are the norm.

\acknowledgements{The Maunakea Spectroscopic Explorer (MSE) conceptual design phase was conducted by the MSE Project Office, which is hosted by the Canada-France-Hawaii Telescope (CFHT). MSE partner organizations in Canada, France, Hawaii, Australia, China, India, and Spain all contributed to the conceptual design. The authors and the MSE collaboration recognize the cultural importance of the summit of Maunakea to a broad cross section of the Native Hawaiian community.}

\bibliography{report} 
\bibliographystyle{spiebib} 

\end{document}